\documentclass{IOS-Book-Article}

\usepackage{mathptmx}
\usepackage[hyphens]{url}
\usepackage{graphicx}
\usepackage[caption=false]{subfig}

\usepackage[linesnumbered,ruled]{algorithm2e}
\usepackage{multirow}
\let\oldnl\nl
\newcommand{\nonl}{\renewcommand{\nl}{\let\nl\oldnl}}
\usepackage{algpseudocode}
\algnewcommand{\And}{\textbf{ and }}

\def\hb{\hbox to 10.7 cm{}}

\begin{document}

\pagestyle{headings}
\def\thepage{}

\begin{frontmatter}              

\title{Edge-as-a-Service:\\
Towards Distributed Cloud Architectures}

\markboth{}{October 2017\hb}

\author[A]{\fnms{Blesson} \snm{Varghese}%
\thanks{Corresponding Author. E-mail: varghese@qub.ac.uk; Web: www.blessonv.com}},
\author[A]{\fnms{Nan} \snm{Wang}},
\author[A]{\fnms{Jianyu} \snm{Li}}
and
\author[A]{\fnms{Dimitrios S.} \snm{Nikolopoulos}}

\runningauthor{B. Varghese et al.}
\address[A]{School of Electronics, Electrical Engineering and Computer Science\\ Queen's University Belfast, UK}

\begin{abstract}
We present an Edge-as-a-Service (EaaS) platform for realising distributed cloud architectures and integrating the edge of the network in the computing ecosystem. The EaaS platform is underpinned by (i) a lightweight discovery protocol that identifies edge nodes and make them publicly accessible in a computing environment, and (ii) a scalable resource provisioning mechanism for offloading workloads from the cloud on to the edge for servicing multiple user requests. We validate the feasibility of EaaS on an online game use-case to highlight the improvement in the QoS of the application hosted on our cloud-edge platform. On this platform we demonstrate (i) low overheads of less than 6\%, (ii) reduced data traffic to the cloud by up to 95\% and (iii) minimised application latency between 40\%-60\%. 
\end{abstract}

\end{frontmatter}
\markboth{July 2017\hb}{July 2017\hb}

\section*{Introduction}
\label{sec:introduction}
The upcoming Internet-of-Things paradigm motivates the need for developing distributed cloud architectures to meet future computing challenges~\cite{paper0}. 
Distributed cloud architectures tap into spare computing that may be available at the edge of the network (for example on local routers, mobile base stations and switches~\cite{meurisch2015upgrading}). 
However, this is challenging and not an easy task. Alternatively, it is proposed that low cost and low power computing nodes, such as Raspberry Pis can be placed one hop away from user devices, sometimes referred to as micro clouds. These nodes can host servers offloaded from cloud servers to service a set of user devices for improving the overall QoS~\cite{enorm}.  

The edge of the network will need to be integrated into the computing ecosystem for realising a distributed cloud architecture. For example, micro clouds will need to be accessible to applications and application owners if a server is to be deployed on them. There are public services to enable the selection of cloud resources and deployment of applications, such as the Amazon Web Services for the Elastic Compute Cloud. However, there are no such services to enable the selection and deployment of applications on edge nodes. This paper addresses the problem by proposing and developing the first \textit{`Edge-as-a-Service'} (EaaS) platform to makes edge nodes publicly available.  

The EaaS platform is built on a discovery protocol. A collection of homogeneous edge nodes are identified and made publicly accessible via a controller. We used Raspberry Pis\footnote{\url{https://www.raspberrypi.org/}} as edge nodes and implemented management algorithms to make edge node services accessible in the platform. The feasibility of the EaaS platform was explored in two ways. Firstly, by experimentally investigating the overheads in the platform. The key observation is that our platform has an overhead of up to 6\%. Secondly, by developing a real use-case, which is a location-aware online game that make use of distributed cloud architectures. The key result is that the application latency is reduced between 40\%-60\% for up to 1024 users and the edge node processes nearly 95\% data. 

\section{Edge-as-a-Service (EaaS) Platform}
\label{sec:eaas}

The EaaS platform operates as a three tier architecture as follows. The top tier is the \textit{cloud layer} which hosts application servers. A centralised cloud architecture will be typically a two-tier model in which user devices connect to the application server. 

The bottom layer is the \textit{device layer} comprising user devices, such as smartphones, wearables and gadgets that connect to cloud application servers. Typically in a centralised cloud architecture, devices connect to the application servers through traffic routing nodes. However, to include computing services offered by edge nodes, user devices will need to connect to edge nodes through the cloud server. 

The middle tier is the \textit{edge node layer} in which edge node(s) are made available on-demand to support a collection of user devices that may be close to the node(s). For example, when users start an application, a connection is firstly established with the cloud application servers. Then an application server (either a clone or a partitioned server) may be deployed on an edge node. 

The EaaS platform is underpinned by mechanisms for (i) discovering edge nodes, and (ii) provisioning resources on edge nodes for workloads.

\subsection{Component View}
\label{sec:component}
\begin{figure}[t]
	\centering
	\includegraphics[width=0.6\textwidth]{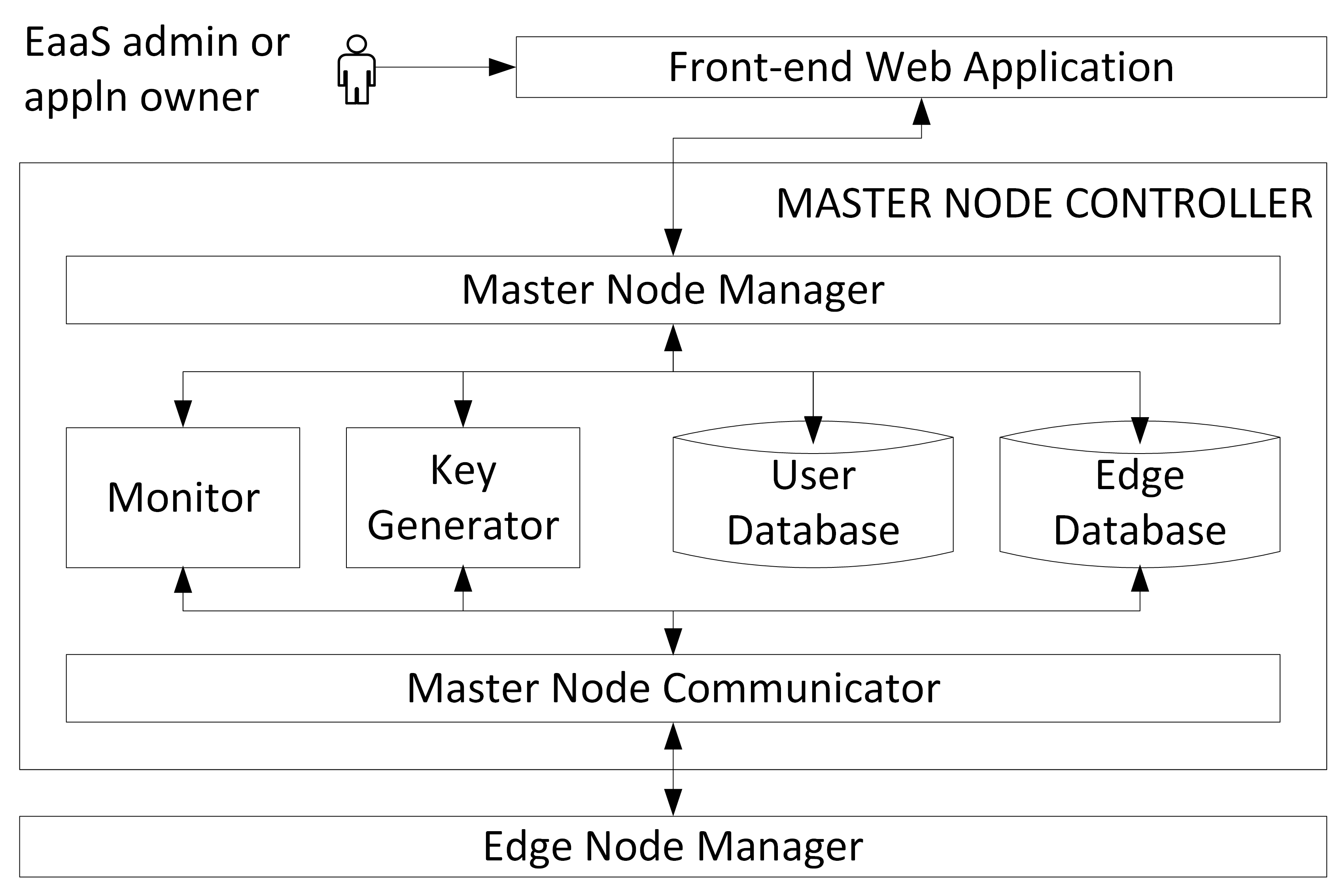}
	\caption{Component view of the EaaS platform}
	\label{fig:figure2}
\end{figure}

The EaaS platform that operates in the edge node layer requires a master node (located either in the edge node layer or elsewhere) and a collection of edge nodes. The master node executes an EaaS controller that connects to a collection of edge nodes. Each edge node executes an EaaS manager. The Master Node Controller comprises the following six modules (refer Figure~\ref{fig:figure2}).

1) The \textit{Master Node Manager} is the interface between the user input from the Web Application and the Master Controller. This module interacts with the Monitor and the Key Generator. Information to isolate users is maintained by the User Database on the Master Node. The Edge Database contains information on all edge nodes that have been discovered and have installed an Edge Manager on the edge node.

2) The \textit{Monitor} periodically obtains key metrics, such as the CPU and memory utilisation from each edge node. The frequency of obtaining this information can be set by the Administrator. However, currently this module is centralised and we anticipate that it will be less scalable. Alternate distributed monitoring mechanisms will need to be implemented to make the EaaS platform more scalable.

3) The \textit{Key Generator} enhances security of the EaaS platform by generating a public and private key for containers. Users are provided can download a private key via the Web Application when a container is launched for the first time. 

4) The \textit{User Database} maintains the users of the EaaS system by managing usernames and passwords. The current active users on the edge nodes and their time of activity are logged in this database. 

5) The \textit{Edge Database} stores the information of all discovered edge node, such as its public IP, open port number and coordinator information and of containers on the edge node, namely its name, state and public IP.

6) The \textit{Master Node Communicator} is the interface between the Controller and the Edge Node and communicates with edge node manager. Commands to launch, start, stop and terminate containers on the edge nodes are assembled in this module. The results of executing the commands on the edge node are obtained from the Edge Node Manager and used to update the database. Additionally, the metric information obtained from Monitoring is obtained by the Monitor via this module.

The \textit{Edge Node Manager} is installed on the edge node during discovery and executes commands obtained from Master Node Communicator and provides the output of the execution back to the Communicator. 

The \textit{Web Application} provides a user interface for accessing the EaaS platform. The users may be administrators of the EaaS platform or application owners who require access to an edge node. The administrator interface provides global information on a variety of metrics, such as CPU utilisation and memory utilisation of containers that are deployed by users on the edge node. The administrator can start, stop or terminate containers that are running. The application owner interface provides information relevant to the containers that a user has deployed on an edge node. The users can choose to launch an application container (for example, Docker) or an Operating System container (for example, LXD) on the edge node.

\subsection{Discovery and Provisioning Services}
\label{sec:services}
\begin{table*}[ht]
	\centering
	\caption{Notation used in the proposed EaaS platform}
	\begin{tabular}{c p{8cm} p{2cm}} \hline
		\textbf{Parameter} & \textbf{Description} & \textbf{Source} \\ \hline
		\(Prt_{controller}\) & Port used on controller for communication with edge nodes & \multirow{ 4}{*}{Master Node Controller}\\ 
		\(IP_{controller}\) & IP address of the controller & \\ 
		\(E\) & A set of $n$ edge nodes offered as a service, $e_i \in E, i=1,..., n$, where $e_i$ = [$IP_{e_i}$] & \\
		\(C\) & A set of $m$ containers launched on $E$, $c_i \in C, i=1,..., m$, where $c_i$ = [$IP_{c_i}, status_{c_i}$] & \\
		\hline
		\(IP_{e_i}\) & IP address of edge node $e_i$ & \multirow{ 8}{*}{Edge Node}\\
		\(Prt_e\) & Port used on edge nodes for communication with controller & \\
		\(offer_i\) & A list of $[e_i,IP_{e_i},Prt_e]$ to offer services on edge node $e_i$ & \\
		\(service\) & Flag on the availability of the services on an edge node & \\
		\(key_{c_i}\) & Private key generated for accessing $c_i$ & \\
		\(result_i\) & Results generated by an application in $c_i$ & \\
		\(IP_{c_i}\) & IP address of container $c_i$ & \\
		\(status_{c_i}\) & Status of container $c_i$ & \\
		\hline
		\(conType\) & Flag for whether an OS or application container is deployed & \multirow{ 3}{*}{Edge Service User}\\
		\(action_i\) & Actions (launch, terminate, start or stop) on a container $c_i$  & \\
		\(request_i\) & A list of $[conType, e_i, action_i, c_i]$ to request services on edge node $e_i$ & \\
		
		\hline
	\end{tabular}
	\label{tab:table1}
\end{table*}

Table~\ref{tab:table1} shows the mathematical notation employed for describing the discovery protocol and the services offered by the EaaS platform. The parameters described in the table are provided either by the Master Node Controller, the Edge Node or the Edge User. The discovery protocol identifies edge nodes that communicate with the master node. Each edge node installs a simple Manager initially in the discovery. 

Algorithm~\ref{algo:discovery} presents the discovery protocol in which edge nodes are brought in a single environment of the EaaS platform. The algorithm shows the activities on the Master Node Controller and the Edge Node Manager. The port to communicate with the Master Node Controller is known to the Edge Node Manager (Line 1). A list of all edge nodes and their services are maintained by the Controller (Lines 2 and 3) because the Edge Node Manager updates the Controller when it can offer a service (Lines 5 and 6).

\begin{algorithm}
\SetAlgorithmName{Procedure}{}
\DontPrintSemicolon
 \nonl on the \textbf{Master Node Controller}\\
 \KwData{$Prt_{controller}, E, offer_i$}
 listen $Prt_{controller}$\;
 \If{$offer_i$}{
   		update $E$ with $e_i$\;}
 \nonl\;
 \nonl on the \textbf{Edge Node Manager}\\
 \KwData{$IP_{controller}, Prt_{controller}, service, offer_i$}
 \If{$service == True$}{
   		send $offer_i$ to controller using $IP_{controller}, Prt_{controller}$\;}
 \caption{Discovery protocol}
 \label{algo:discovery}
\end{algorithm}

Procedure~\ref{algo:serviceonfrontend}, Procedure~\ref{algo:serviceoncontroller} and Procedure~\ref{algo:serviceonedgenodemanager} presents how different services are offered in the EaaS platform and the corresponding functions on the Front-end, Master Node Controller and Edge Node Manager respectively. 

On the Front-end, a user requests services to the Controller for either launching, terminating, starting or stopping containers on an edge node (Line 1). If the request is to start, stop or terminate containers then these actions are performed. However, if an OS container is to be launched then the user is provided access to the container via a private key (Lines 3-6). If an application container is to be launched, then it is executed on the edge node and the results are obtained (Line 9).

\begin{algorithm}
\SetAlgorithmName{Procedure}{}
\DontPrintSemicolon
 \KwData{$Prt_{controller}, IP_{controller}, request_i, key_{c_i}, result_i$}
 send $request_i$ to controller using $IP_{controller}, Prt_{controller}$\;
 \If{$action$ == 'launch'}{
 \eIf{$conType$ == 'os'}{
   \If{$key_{c_i}, IP_{c_i}$}{
   		download $key_{c_i}$\;
        access to $IP_{c_i}$\;
   		}
   }{
   receive $result_i$\;
   }}
 \caption{Edge service procedure on the \textbf{Front-End}}
 \label{algo:serviceonfrontend}
\end{algorithm} 

On the Master Node Controller, when a request is received from the Front-End it is passed on to the Edge Node Manager (Line 1). If the request is to launch an OS container then the private key generated by the Controller is provided to the user (Lines 3-4). The list of containers on each edge node is updated (Line 6). Alternatively, if the request from the Front-end was to launch an application container, then the results obtained from the Edge Node Manager are displayed on the Front-end (Line 9). 

\begin{algorithm}
\DontPrintSemicolon
\SetAlgorithmName{Procedure}{}
 \KwData{$request_i, C, key_{c_i}, IP_{c_i}, status_{c_i}, result_i$}
 send $request_i$ to $e_i$\;
 \eIf{$conType$ == 'os'}{
 	\If{$key_{c_i}, IP_{c_i}$}{
   		send $key_{c_i}, IP_{c_i}$ to Front-end\;}
 	update $C$ with $status_{c_i}, IP_{c_i}$\;}
    {\If{$result_i$}{
    	send $result_i$ to Front-end\;}}
 \caption{Edge service procedure on the \textbf{Master Node Controller}}
 \label{algo:serviceoncontroller}
\end{algorithm}

On the Edge Node Manager, LXD commands are launched for an OS container and Docker containers are used for application containers. If the request is to launch a container, then Lines 4-7 are executed. If an application container is to be launched then lines 9-10 are executed. Similarly for start (Lines 14-16), for stop (Lines 19-20) and for terminate (Lines 23-24) are executed. 

\begin{algorithm}
\DontPrintSemicolon
\SetAlgorithmName{Procedure}{}
 \KwData{$request_i$}
 \Switch{action}{
 	\Case{launch}{
    	\eIf{$conType$ == 'os'}{
   		launch LXD container $c_i$\;
        generate $key_{c_i}$\;
        configure $IP_{c_i}$\;
        send $key_{c_i}, IP_{c_i}, status_{c_i}$ to controller\;
   		}{
       	launch Docker container $c_i$\;
        send $result_i$ to controller\;}}
    \Case{start}{
     	start LXD container $c_i$\;
        configure $IP_{c_i}$\;
        send $IP_{c_i}, status_{c_i}$ to controller\;}
    \Case{stop}{
    	stop LXD container $c_i$\;
        send $status_{c_i}$ to controller\;}
    \Case{terminate}{
    	terminate LXD container $c_i$\;
        send $status_{c_i}$ to controller\;}
    }
 \caption{Edge service procedure on the \textbf{Edge Node Manager}}
 \label{algo:serviceonedgenodemanager}
\end{algorithm}

\section{Use-case}
\label{sec:usecase}
The feasibility of EaaS is demonstrated on an open-sourced version of a location-aware online game similar to Pok\'eMon Go, named iPokeMon\footnote{\url{https://github.com/Kjuly/iPokeMon}}. The game features a virtual reality environment that can be played on a variety of devices, such as smartphones and tablets. The user locates, captures, battles and trains virtual reality creatures, named Pok\'emons, through the GPS capability of the device. The Pok\'emons are geographically distributed and a user aims to build a high value profile among their peers. The users may choose to walk or jog through a city to collect Pok\'emons. 

The current execution model uses a centralised cloud architecture, such that the game server is hosted on the public cloud and the users connect to the server. The server updates the user position and a global view of each user and the Pok\'emons is maintained by the server. For example, if Amazon Elastic Compute Cloud (EC2) servers are employed, then the game may be hosted in an EC2 data center and a user in Belfast communicates with the game server. The original game server is known to have crashed multiple times during its launch due to severe activities which were not catered for\footnote{\url{http://www.forbes.com/sites/davidthier/2016/07/07/pokemon-go-servers-seem-to-be-struggling/\#588a88b64958}}.

\begin{figure}
	\centering
	\includegraphics[width=0.8\textwidth]{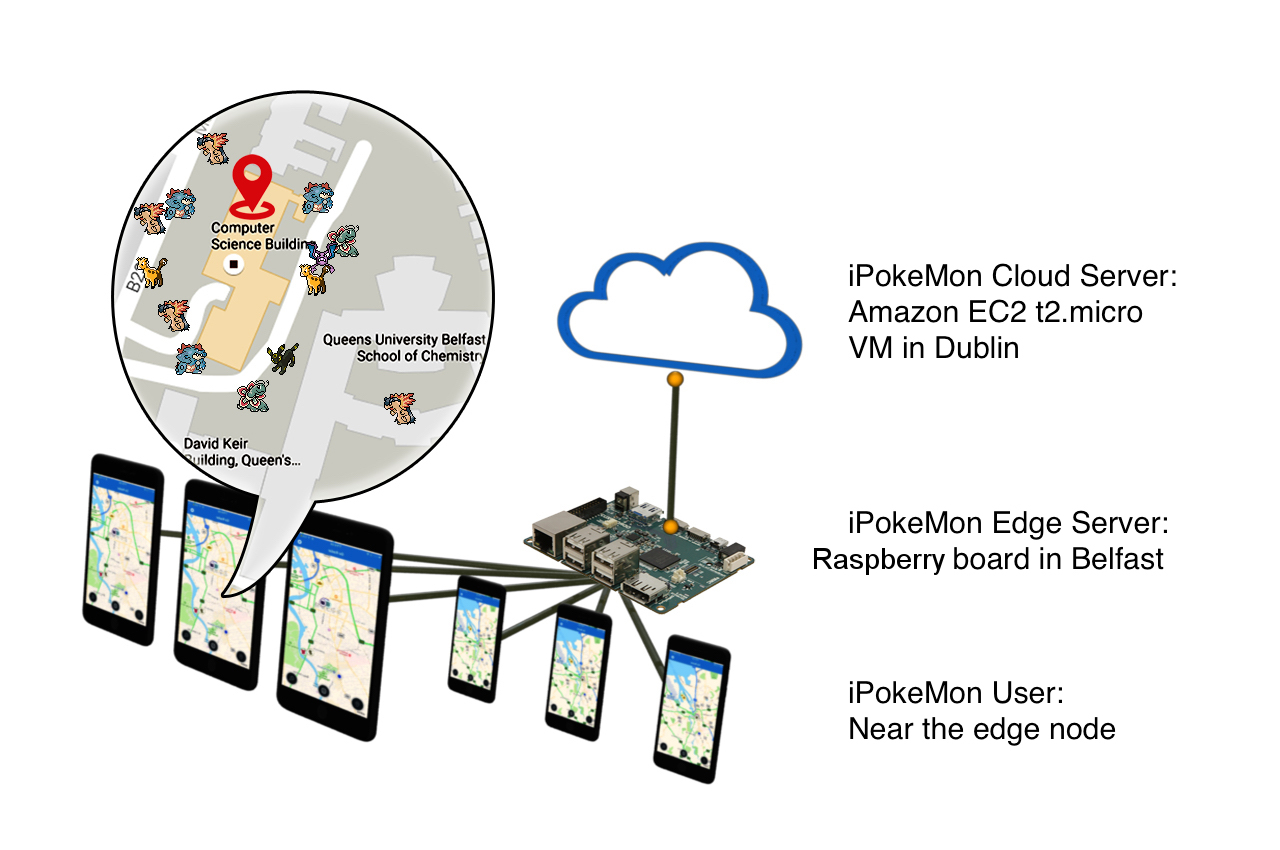}
	\caption{Distributed iPokeMon game using the EaaS plaform.}
	\label{fig:edgeBasediPokeMon}
\end{figure} 


We used the EaaS platform to employ a distributed cloud architecture for executing the iPokeMon game as shown in Figure~\ref{fig:edgeBasediPokeMon}. The game server is hosted in the Amazon EC2 Dublin data center on a t2.micro instance. A partitioned game server is deployed using an LXD container on an edge node which executes the edge node manager of the EaaS platform. The server was manually partitioned (the cloud server maintained a global view of the Pok\'emons, where as the edge node server had a local view of the users connected to the edge server). The edge node periodically updated the global view of the cloud server. User devices connect to the edge node instead of the cloud server to service requests (process GPS coordinates and update personal profile). 

\section{Evaluation}
\label{sec:evaluation}
In this section, we evaluate the proposed EaaS platform by measuring the overheads associated with deploying workloads on edge nodes. Before this we consider the test-bed we have developed for the EaaS platform.

\subsection{Platform}
The edge nodes used are a set of three Raspberry Pi 2 boards. Each board comprises a 900MHz quad-core ARM Cortex-A7 CPU and 1 GB RAM. An Ubuntu MATE 16.04.2 LTS image is copied to a micro SD card that is used by the Raspberry Pi as the operating system. LXD containers are employed for launching workloads on the Raspberry Pi. The Operating System (OS) of the containers that are launched is Alpine Linux.
This OS creates a lean container (other OS' have additional packages installed). The Edge Node Manager is executed locally on the Pis. The edge nodes are connected via a switch to an EaaS master node. The master node is an Intel i7-4720HQ CPU @ 2.60GHz 2.59GHz and 8 GB RAM system. 

\subsection{Measuring Overheads}


\begin{figure}
	\centering
	\includegraphics[width=0.58\textwidth]
	{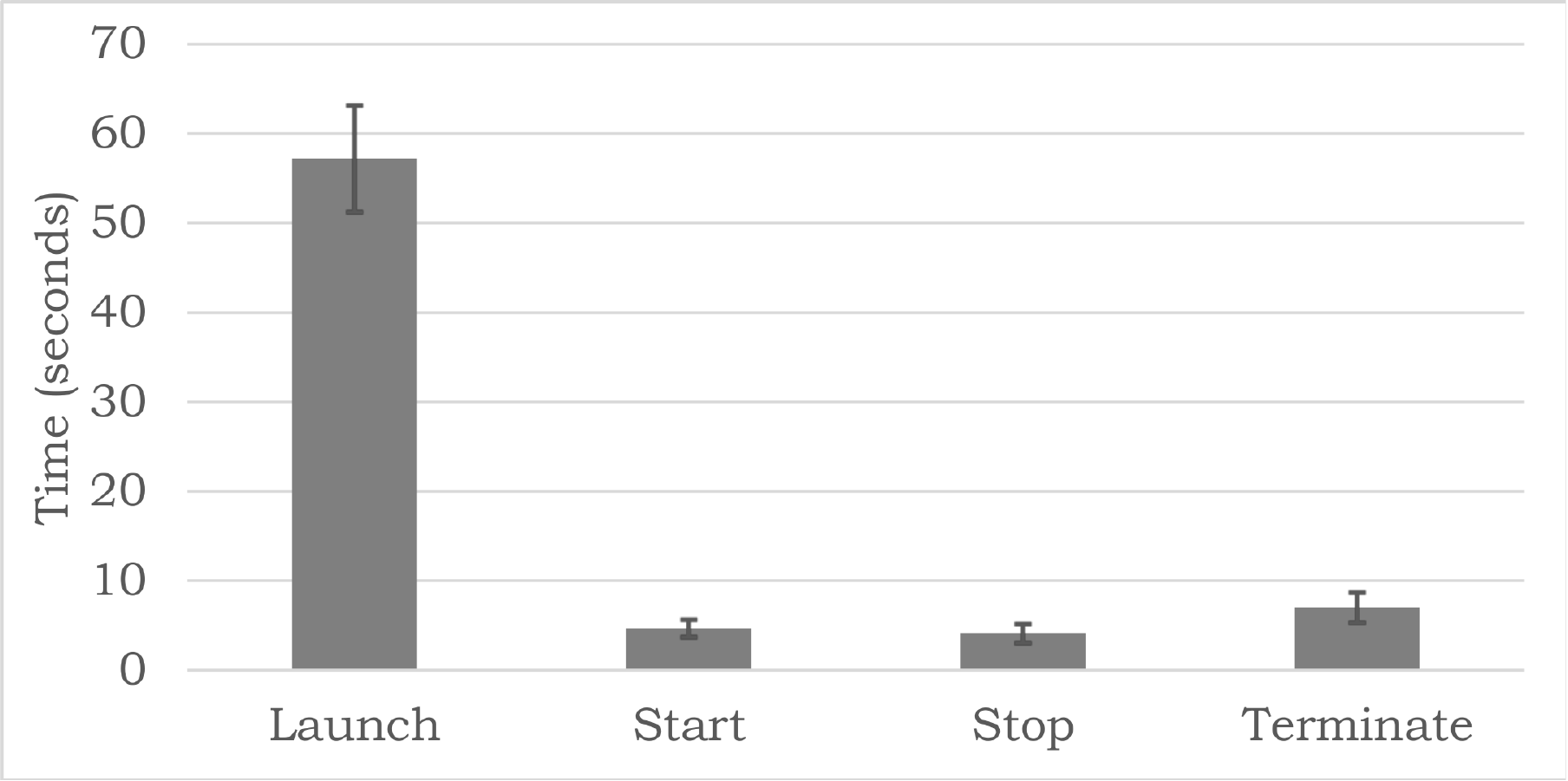}
	\caption{Overall round trip latency}
	\label{figure4}
\end{figure}

The average overhead involved in using the EaaS platform on three edge nodes of our experimental test-bed was considered. 


Our experiments indicated that the overheads remain nearly constant when 50 containers are used (these results are exhaustive and not in the scope of this paper). The time taken to launch a container is under 1 minute on each edge node, even when 50 containers are launched. A container can be started and stopped in under 5 seconds and terminated in nearly 7 seconds. The results show that the EaaS platform is scalable up to 50 containers. Experiments were not pursued beyond 50 containers since the resources are exhausted on the edge node by the container. 

Figure~\ref{figure4} shows the average overall round trip latency (when the user initiates the operation on the front-end, which is then passed on to the master node and after which the edge node manager executes the operation on the edge node) for launching, starting, stopping and terminating containers and the standard deviation on the EaaS platform. The average overhead that is added by the front-end and master node in the EaaS platform for managing the edge nodes is shown in Table~\ref{tab:table2}. 
We do not account for the time taken for launching, starting, stopping and terminating a container on an edge node in calculating the overhead since it cannot be eliminated if containers are used on the edge node. 
An overall overhead of nearly 6\% is added by the EaaS platform for launching the container. This is higher than the overheads for starting, stopping and terminating containers because the master node deals with generating the key on the edge node. 

\begin{table}[ht]
	\caption{Percentage overhead in using the front-end and master node of the EaaS platform}
	\label{tab:table2}
	\begin{center}
		\centering    
		\begin{tabular}{l p{2.8cm} p{3cm} p{1.2cm}}
			\hline	
			\multirow{2}{*}{\textbf{Service}} &	\multicolumn{3}{c}{\textbf{Communication Overhead (\%)}}\\
			\cline{2-4}
			&	Between front-end and EaaS master node	&	Between EaaS master node and an edge node & Overall\\
			\hline
			\textbf{Launch} &	0.54	&	5.36	&	5.90\\
			\textbf{Start} & 0.06	&	0.91	&	0.97\\
			\textbf{Stop} & 0.07	&	1.06	&	1.13\\
			\textbf{Terminate} & 0.04	&	0.64	&	0.68\\
			
			\hline 		
		\end{tabular}%
	\end{center}
\end{table}

\subsection{EaaS for the Use-case}

\begin{figure*}
	\centering
	\subfloat[Latency of iPokeMon game users when using a server located on the cloud and on an edge node.]
	{	\label{fig:usecase1-1}
		\includegraphics[width=0.48\textwidth]
		{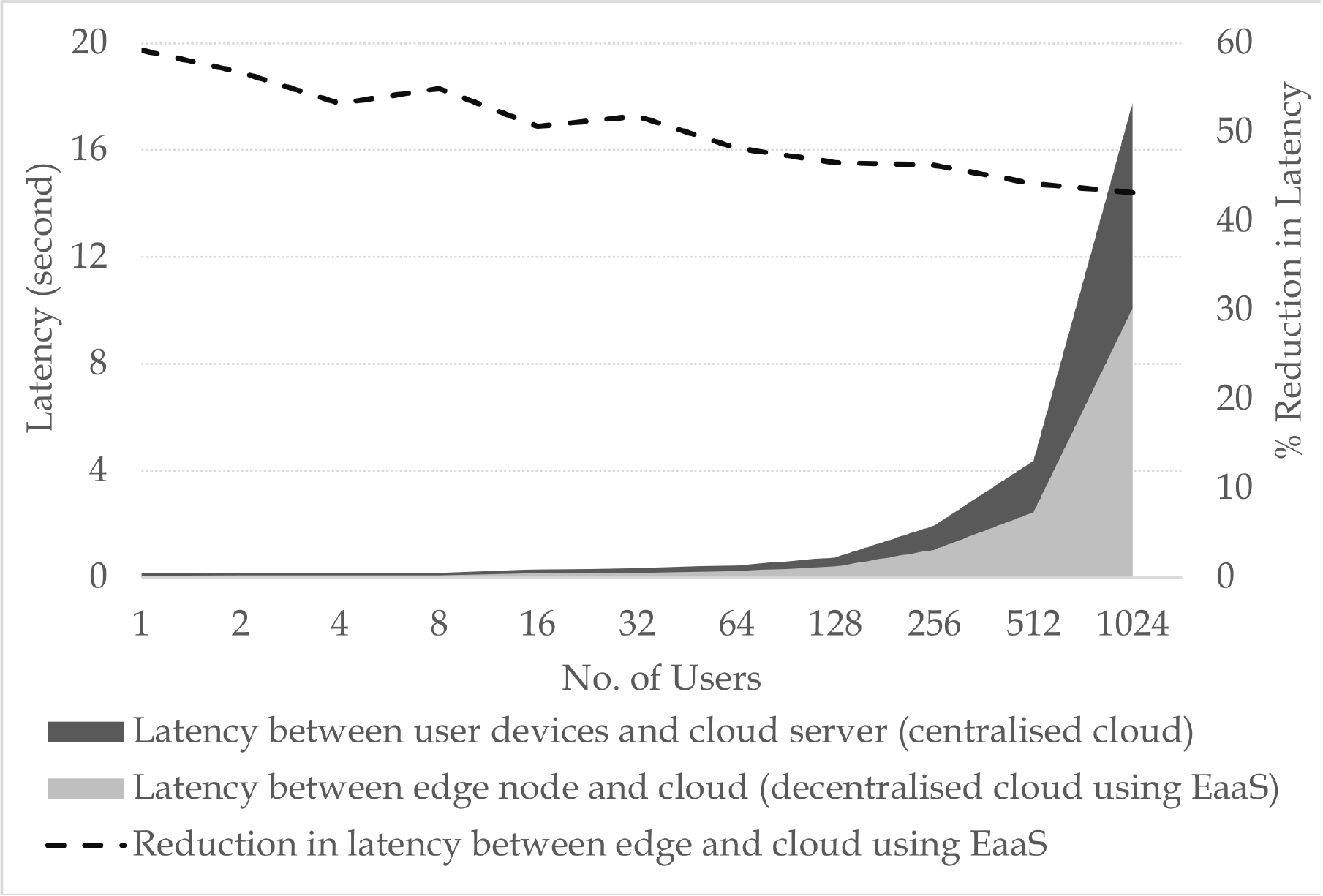}
	}
	\hfill
	\subfloat[Percentage reduction in the data traffic between edge nodes and the cloud.]
	{	\label{fig:usecase1-2}
		\includegraphics[width=0.485\textwidth]
		{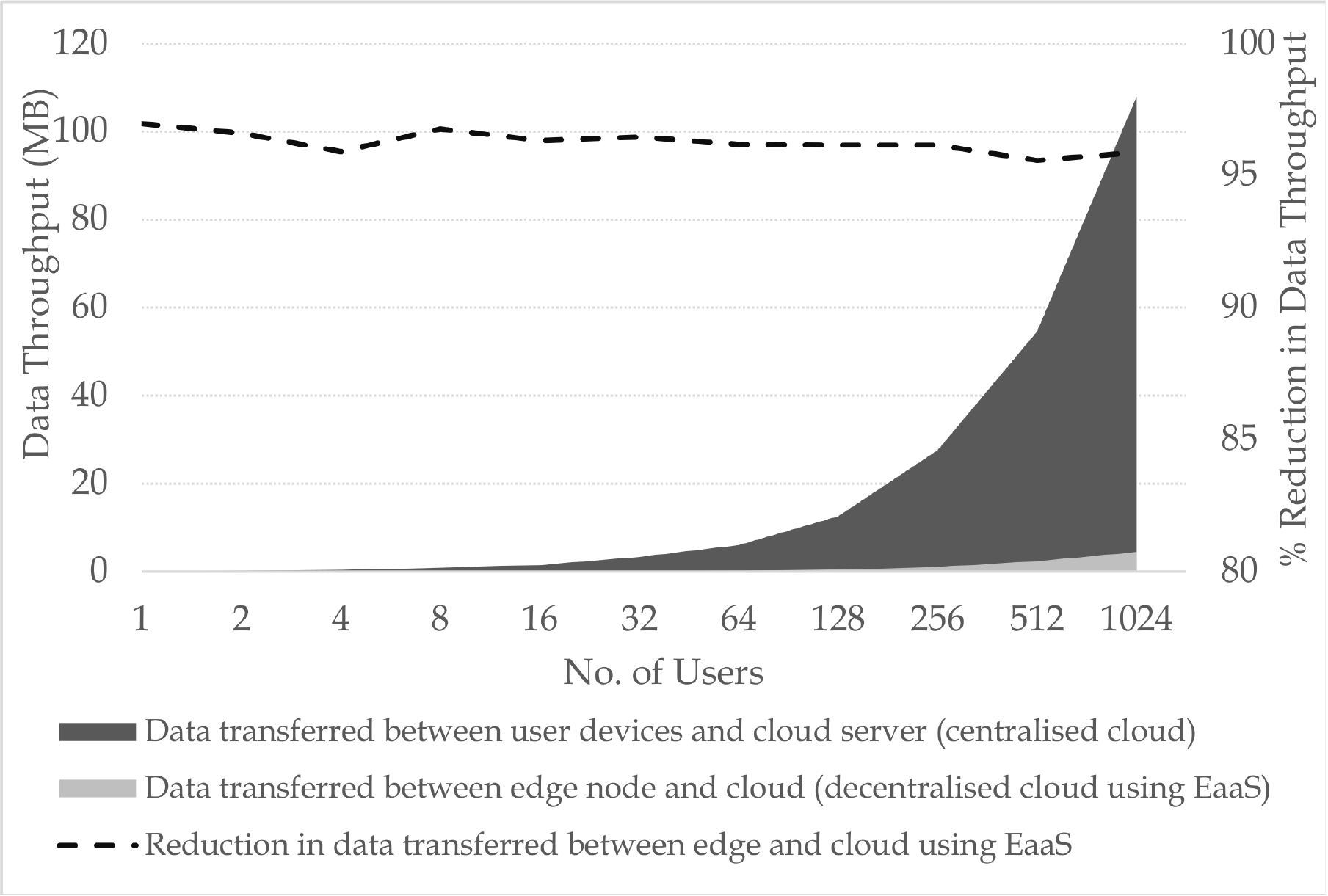}
	}
\caption{Latency and reduction in data traffic using EaaS}	
\end{figure*}

Figure~\ref{fig:usecase1-1} shows the average latency experienced by the user, which is measured by round trip latency from when the user device generates a request while playing the game that needs to be serviced by a cloud server (this includes the computation time on the server). The response time is noted over a five minute time period for varying number of users. Using EaaS for a decentralised cloud architecture, it is noted that on an average the latency is reduced between 40\%-60\% for up to 1024 users playing the game. 

Figure~\ref{fig:usecase1-2} presents the amount of data that is transferred during a fifteen minute time period. As expected with increasing number of users the data transferred increases. However, we observe that using EaaS the data transferred between the edge node and the cloud is significantly reduced, yielding an average of over 95\% reduction.

\section{Related Work}
\label{sec:relatedwork}
Different architectures have been proposed at the data center level for achieving distributed clouds. For example, 
hybrid~\cite{hybridcloud-1},  federated~\cite{federatedcloud-1} and ad hoc~\cite{adhoccomputing-1} clouds. 

More recently to address the connectivity and data challenges that are expected from the IoT paradigm cloud architectures extending to resources outside data centers have been proposed~\cite{iot-1}. They decentralise the concentration of compute resources away from data centers towards the edge of the network~\cite{microcloud-0}. This helps in bringing computing closer to the source of data thereby reducing traffic beyond the first hop in the network. 

The deployment of cloudlets, which are dedicated nodes at the edge of the network can service requests of mobile devices in a distributed manner~\cite{cloudlet-1,cloudlet-2}. 
Hence, mobile edge computing has emerged~\cite{mec-1}.
Similarly, micro clouds, which are miniature data centers are reported in literature~\cite{microcloud-2,microcloud-4}.
This has resulted in edge and fog computing~\cite{offload-1,offload-3}.

Despite the advent of these architectures it is challenging to integrate the edge of the network in a cloud architecture~\cite{osmotic-1}. The key problem that needs to be addressed is making edge nodes publicly accessible. This will allow application servers running on the cloud to offload workloads onto edge nodes to service requests originating from a given location. To this end, edge nodes will need to be firstly discovered and then be brought under a single environment that makes them visible. This paper aims to address this gap by developing an Edge-as-a-Service platform. 

\section{Conclusions}
\label{sec:conclusions}
The EaaS we propose and develop aims to integrate the edge of the network in the computing ecosystem. This paper presents the first steps we have taken in realising distributed cloud architectures for meeting the challenges of emerging distributed workloads.

\bibliographystyle{unsrt}
\bibliography{references1} 

\end{document}